\documentclass[12pt]{article}

\usepackage{epsfig}
\usepackage{graphicx}
\usepackage{a4}
\usepackage{latexsym}
\usepackage{cite}

\usepackage{pslatex}
\usepackage[latin1]{inputenc}
\usepackage[T1]{fontenc}
\newcommand{\ep}{\mbox{$\varepsilon$}}

\begin{document}
\setlength{\parskip}{0.2cm}
\setlength{\baselineskip}{0.55cm}
\begin{titlepage}
\begin{flushright}
\hfill {\tt hep-ph/xxxxxxx}
\\
\hfill { HRI-04/2006}
\\ \hfill {DESY 06-054}
\\ \hfill {SFB/CPP-06-22}
\end{flushright}
\vspace{5mm}

\vspace{2cm}
\begin{center}
{\Large \bf
\boldmath
QCD threshold corrections to Higgs decay and to hadroproduction
in $l^+l^-$ annihilation}\\
\end{center}

\vspace{2cm}
\begin{center}
{\bf
J.~Bl\"umlein$^a$  and
V.~Ravindran$^{b}$}
\end{center}
\begin{center}
{\it $^a$~Deutsches Elektronen--Synchrotron, DESY,\\
Platanenallee 6, D--15738 Zeuthen, Germany}
\\
\vspace{2mm}
{\it $^b$~Harish--Chandra Research Institute, Chhatnag Road,\\
 Jhunsi, Allahabad, India.}
\end{center}

\vspace{4cm}
\begin{center}
{\bf ABSTRACT}
\end{center}
We present threshold enhanced QCD corrections to the bottom quark energy 
distribution in Higgs boson decay and to hadroproduction in $l^+l^-$ 
annihilation beyond leading order in the strong coupling constant.  This 
is achieved using the resummed decay distribution obtained using 
renormalisation group invariance and the mass factorisation theorem that 
they satisfy and Sudakov resummation of soft gluons.
\vskip12pt
\vskip 0.3 cm

\end{titlepage}
The Higgs boson (H) which emerges in the
electroweak symmetry breaking mechanism
\cite{Djouadi:2005gi}
is the only particle that is yet to be discovered in the Standard Model(SM).
Searches at LEP experiments \cite{unknown:2005em} indicate that the
lower bound on its mass is around $114.2~GeV$ and
the upper bound around $260~GeV$ at $95\%$ confidence level.
The hadronic machine Tevatron at Fermi-lab currently running
with increased energy and the upcoming Large Hadron Collider (LHC)
at CERN have the physics goal
of discovering the Higgs boson.  At hadronic colliders,
among the potential channels through which the Higgs boson
can be produced, the vector boson fusion process is one of the
promising ones \cite{Buttar:2006zd}.
Here the Higgs boson is produced in
a reaction $P_1+P_2 \rightarrow V+H$, where $P_i$ are
the incoming hadrons, $V$ is
the vector boson $W^\pm/Z$, followed by
the dominant decay mode $H \rightarrow b +\overline b$.  For the
heavy Higgs boson searches, the decay to bottom quarks can be
accessible in the Two-Higgs Doublet models due to enhanced couplings
\cite{Djouadi:2005gj}.

A similar process in nature is inclusive hadroproduction in
$l^+ l^-$ annihilation \cite{Altarelli:1979kv}.
Here the incoming leptons annihilate into
vector boson such as $\gamma^*/Z$ which then fragments into hadrons,
i.e., $l^+ l^- \rightarrow \gamma^*,Z\rightarrow P+X,$ where
$P$ is any hadron and $X$ is the remaining final state.
The double differential cross section for producing
a hadron with energy fraction $x$ of the parton produced
in the annihilation is given by
\begin{eqnarray}
{d^2 \sigma^P \over dx ~d\cos\theta}
&=& {3 \over 8} (1+ \cos^2 \theta)~ {d \sigma^P_T \over dx}(x,q^2)
+{3 \over 4} \sin^2 \theta~{d \sigma^P_L\over dx}(x,q^2)
+{3 \over 4} \cos \theta~ {d \sigma^P_A \over dx}(x,q^2)~,
\end{eqnarray}
where $\theta$ is the center of mass angle of the final state hadron.
The energy fraction is given by $x=2 p\cdot q/q^2$, where $p$ and $q$ are
the momenta of the final state hadron and
the intermediate vector boson respectively.
The $B$ hadron production in $l^+l^-$ channel is an important
process to determine the non-perturbative $b$ quark fragmentation functions
\cite{Corcella:2005dk}.
Precise knowledge of them will reduce theory uncertainties
in Higgs decays to bottom quarks and also in the top quark mass
reconstructions in the top quark decays at Tevatron and LHC.

In the case of the decay process $H\rightarrow b+\overline b$, we
study the energy distribution of one of the bottom quarks.
We use the perturbative fragmentation approach \cite{Mele:1990cw}
which is
valid when $m_b \ll m_H$ to factorise the
decay distribution into a part which contains
the Higgs boson decay to massless partons and a part containing perturbative
fragmentation functions denoted by $D_I^b$ that describe the fragmentation
of massless partons into massive bottom quarks.   In this approach, the large
logarithms $\log(m_H^2/m_b^2)$, where $m_H$ is the Higgs boson mass
and $m_b$ bottom quark mass, can be resummed using
perturbative fragmentation functions.  The normalised decay distribution 
is given by
\begin{eqnarray}
{1 \over \Gamma_0} {d \Gamma_b \over d x_b}(x_b,q^2,m_b^2) &=& \sum_I
\int_{x_b}^1 {d z\over z} ~
C_I\left(z,q^2,\mu_R^2,\mu_F^2\right)
D_I^b\left({x_b \over z},m_b^2,\mu_F^2\right) \quad \quad \quad I=b,g,
\end{eqnarray}
where $\Gamma_0$ is the Born decay distribution, $x_b=2 p_b\cdot q/q^2$ 
with $p_b,q$ the momenta of $b$ quark and the Higgs boson, with
masses $m_b$ and $m_H$ respectively.  Here $q^2=m_H^2$.
The renormalisation scale $\mu_R$ is due to
ultraviolet renormalisation and the factorisation scale $\mu_F$ is due to
mass factorisation of the collinear singularities.
The fixed order result for $C^b_I$ for $I=b$ up to next to leading order 
(NLO) in strong
coupling constant $\alpha_s$ is available in the literature \cite{Corcella:2004xv}.
The soft gluons that result from the outgoing partons
lead to large logarithms that can be resummed systematically.  The impact
of such effects up to the next to leading logarithmic (NLL) level has been
studied along with fixed order NLO QCD corrections
in \cite{Corcella:2004xv,Corcella:2005dk}.

The inclusive hadroproduction in $l^+l^-$ annihilation process
factorises as
\begin{eqnarray}
{1\over \sigma^{(0)}}{d \sigma_T^P(x,q^2) \over dx}=\sum_{I} \int_x^1 {dz \over z}
C_I(z,q^2,\mu_R^2,\mu_F^2)
D_I^P\left({x \over z},\mu_F^2\right) \quad \quad \quad I=q,\overline q, g
\end{eqnarray}
where $z=2p_I\cdot q/q^2$ is the partonic energy fraction.  Here,
$p_I$ is the momentum of the parton that fragments into the final
state hadron with momentum $p$.  The non-perturbative
function $D_I^P(x/z,\mu_F^2)$ is
the fragmentation function that describes the fragmentation of
the parton of type $I$ into hadron of type $P$.  The perturbatively
calculable coefficient function $C_I$ is known up to NNLO
in QCD for $I=q,\overline q,g$,
\cite{Rijken:1996ns,Rijken:1996np,Rijken:1996vr,Mitov:2006ic}.
The NNLL soft gluon resummation to this process is
also known in the literature
\cite{Kodaira:1981nh,Sterman:1986aj,Catani:1989ne,Vogt:2000ci,Cacciari:2001cw}.

Even though perturbative Quantum Chromodynamics (pQCD) provides a 
framework
to successfully compute various observables involving hadrons at
high energies with less theoretical uncertainty for the physics studies,
the fixed order QCD predictions often have limitations in applicability
due to the presence of various logarithms
that become large in some kinematic regions which otherwise can be
probed by the experiments.  The standard approach to probe these regions is
to resum the class of such large logarithms supplemented with
fixed order results.  This can almost cover the dominant kinematic region of
the phase space.

In this paper, we follow the method that we used in 
Refs.~\cite{Ravindran:2005vv,Ravindran:2006cg} to study the
soft gluon effects.  This generalises our earlier approach
to include any infrared safe decay distribution in perturbative QCD.
We systematically formulate a framework
to resum the dominant soft gluon contributions in $z$ space
to the decay distributions, where $z$ is the appropriate
scaling variable that enters in the process.
We have followed a similar approach described in
\cite{Ravindran:2005vv,Ravindran:2006cg}
that uses renormalisation group (RG) invariance, mass factorisation and
Sudakov resummation of soft gluons as the guiding principles.
Using the resummed results in $z$ space, we compute the soft plus
virtual part of the dominant partonic decay distributions beyond
$NLO$.  In Ref.~\cite{Ravindran:2005vv} we
determined the threshold exponents $D_i^I$ and $B_{i,DIS}$ up to the three
loop level for Drell-Yan process, Higgs boson productions and
deeply inelastic scattering (DIS) cross sections using our resummed
soft as well as soft plus jet distribution functions.
Here we extend the similar all order proof which establishes the relation
between soft plus jet distribution functions
and the threshold resummation exponents and demonstrate the usefulness of this
approach to derive higher order threshold enhanced corrections
for any infrared safe decay distributions.

We find that the soft plus jet distribution relevant for our present study
can be gotten entirely from that of DIS due to the crossing symmetry 
between them
\cite{Drell:1969wb,Suri:1971dc,Gatto:1973sy,Dahmen:1973mn,Furmanski:1981cw}.
In fact, we find that they are identical.  In addition, the Altarelli-Parisi(AP) splitting
functions that determine the scale evolution of the fragmentation functions
coincide with the AP splitting functions of parton distribution functions
in the threshold region.
Extensive discussions on this topic can be found in
\cite{Floratos:1981hs,Stratmann:1996hn,Blumlein:1998hz,Blumlein:2000wh,Dokshitzer:2005bf,Blumlein:2006rr}.
With these
two essential ingredients and using the method described in
\cite{Ravindran:2005vv,Ravindran:2006cg}
we could successfully
reproduce the fixed order NLO soft plus virtual part as well as
the NLL resummation exponents \cite{Corcella:2004xv}
for the bottom quark production in Higgs decay and
also the fixed order NNLO soft plus virtual part
\cite{Rijken:1996ns,Rijken:1996np,Rijken:1996vr,Mitov:2006ic}
and the NNLL resummation exponents
\cite{Kodaira:1981nh,Sterman:1986aj,Catani:1989ne,Vogt:2000ci,Cacciari:2001cw}
for the hadron production in $l^+l^-$ annihilation.
In addition, we can compute the fixed order NNLO
soft plus virtual part of the $b$ quark energy distribution
in the Higgs boson decay and also
partial but dominant $N^iLO$ results for $i=3,4$.
With the same level of accuracy, we can obtain
$N^iLO$ with $i=3,4$ soft plus virtual contributions to hadroproduction in $l^+l^-$ annihilation.

In this article, we will mainly concentrate on
dominant contributions coming from the threshold effects.
This contribution to the
energy distribution of the bottom quark in Higgs boson decay
and to the hadroproduction in the $l^+l^-$ annihilation process
can be obtained by adding the soft part of
the decay distributions with the renormalised virtual corrections and performing
mass factorisation using appropriate mass factorisation counter terms.
We call this infra-red safe combination a "soft plus virtual"($sv$)
part of the decay distribution.
The soft plus virtual part denoted by $\widetilde \Delta^{sv}_{~I}(z,q^2,\mu_R^2,\mu_F^2)$
of the perturbative decay distributions $C_I$
after mass factorisation is found to be
\begin{eqnarray}
{\widetilde \Delta}^{sv}_{~I}(z,q^2,\mu_R^2,\mu_F^2)={\cal C} \exp
\Bigg({\widetilde \Psi^{I}(z,q^2,\mu_R^2,\mu_F^2,\ep)}\Bigg)\Bigg|_{\ep=0}
\label{master}
\end{eqnarray}
where $\widetilde \Psi^I(z,q^2,\mu_R^2,\mu_F^2,\ep)$ is a finite distribution.
For a Higgs boson decaying to bottom quarks, we put $I=b$ and
for that decaying to gluon, $I=g$.  For the hadroproduction in
$l^+l^-$ annihilation, $I=q,\overline q$.
Here $\widetilde \Psi^I(z,q^2,\mu_R^2,\mu_F^2,\ep)$
is computed in $4+\ep$ dimensions, where $n$ denotes the
number of space--time dimensions.
\begin{eqnarray}
\widetilde \Psi^I(z,q^2,\mu_R^2,\mu_F^2,\ep)&=&
\Bigg(
\ln \Big(Z^I(\hat a_s,\mu_R^2,\mu^2,\ep)\Big)^2
+\ln \big|\hat F^I(\hat a_s,Q^2,\mu^2,\ep)\big|^2
\Bigg)
\delta(1-z)
\nonumber\\[2ex]
&&+2~ \widetilde \Phi^{~I}(\hat a_s,q^2,\mu^2,z,\ep)
-{\cal C}\ln \widetilde \Gamma_{II}(\hat a_s,\mu^2,\mu_F^2,z,\ep),
\quad \quad I=b,g,q,\overline q
\label{DYH}
\end{eqnarray}
The symbol "${\cal C}$"
means convolution.  For example, ${\cal C}$ acting on an exponential
of a function $f(z)$ has the following expansion:
\begin{eqnarray}
{\cal C}e^{\displaystyle f(z) }= \delta(1-z)  + {1 \over 1!} f(z)
 +{1 \over 2!} f(z) \otimes f(z) + {1 \over 3!} f(z) \otimes f(z) \otimes f(z)
+ \cdot \cdot \cdot
\end{eqnarray}
The function $f(z)$
is a distribution of the kind $\delta(1-z)$ and ${\cal D}_i$,
\begin{eqnarray}
{\cal D}_i=\Bigg[{\ln^i(1-z) \over (1-z)}\Bigg]_+
\quad \quad \quad i=0,1,\cdot\cdot\cdot
\end{eqnarray}
and the symbol $\otimes$ denotes the Mellin convolution.
We drop all the regular functions that result from
various convolutions.
$\hat F^I(\hat a_s,Q^2,\mu^2,\ep)$ are the form factors
that enter these processes.  For the Higgs boson decay to
bottom quarks, it is related to the Yukawa type of interaction
and to the gluons, it is related to Higgs boson field coupled
to the kinetic term of the gauge fields.  For the $l^+l^-$ annihilation,
the form factor arises from the vector current.
In the form factors, $Q^2=-q^2=-m_H^2$ for the Higgs boson decay and
$Q^2=-q^2=-q_{\gamma^*/Z}^2$.
The functions $\widetilde \Phi^{~I}(\hat a_s,q^2,\mu^2,z,\ep)$ are
called the soft plus jet distribution functions.
The unrenormalised (bare) strong coupling constant
$\hat a_s$ is defined as
$\hat a_s={\hat g^2_s / (16 \pi^2)}$
where $\hat g_s$ is the strong coupling constant which is dimensionless in
$n=4+\ep$.
The scale $\mu$ comes from
the dimensional regularisation in order to make the bare coupling constant $\hat g_s$
dimensionless in $n$ dimensions.
The bare coupling constant $\hat a_s$ is related to the renormalised one 
by
the following relation:
\begin{eqnarray}
S_{\ep} \hat a_s = Z(\mu_R^2) a_s(\mu_R^2) \left(\mu^2 \over \mu_R^2\right)^{\ep \over 2}
\label{renas}
\end{eqnarray}
The scale $\mu_R$ is the
renormalisation scale at which the renormalised strong coupling constant
$a_s(\mu_R^2)$ is defined.  The factorisation scale $\mu_F$ is due
to mass factorisation and the constant
$S_{\ep}=exp\left\{(\gamma_E-\ln 4\pi)\ep/2\right\}$
is the spherical factor characteristic of $n$-dimensional regularisation.
The renormalisation constant up to three loop is given by
\begin{eqnarray}
Z(\mu_R^2)= 1+ a_s(\mu_R^2) {2 \beta_0 \over \ep}
           + a_s^2(\mu_R^2) \Bigg({4 \beta_0^2 \over \ep^2 }+
                  {\beta_1 \over \ep} \Bigg)
           + a_s^3(\mu_R^2) \Bigg( {8 \beta_0^3 \over \ep^3}
                   +{14 \beta_0 \beta_1 \over 3 \ep^2}
                   +{2 \beta_2 \over 3 \ep}\Bigg)
\end{eqnarray}
The renormalisation constant $Z(\mu_R^2)$ relates the bare coupling constant
$\hat a_s$ to the renormalised one $a_s(\mu_R^2)$ through
Eq.~(\ref{renas}).
The coefficients $\beta_0$,$\beta_1$ and $\beta_2$ are
the coefficients that appear in RG equation for the strong coupling constant
up to three loop level\cite{vanRitbergen:1997va,Czakon:2004bu}

The factors $Z^I(\hat a_s,\mu_R^2,\mu^2,\ep)$ are the overall
operator renormalisation constants which renormalise the
bare form factors $\hat F^I(\hat a_s,Q^2,\mu^2,\ep)$.
For the vector current  $Z^q(\hat a_s,\mu_R^2,\mu^2,\ep)=1$,
but for the bottom quarks and gluons coupled to the Higgs boson, the
corresponding form factors
get overall renormalisations 
\cite{vanRitbergen:1997va,Chetyrkin:1997un,GRAC}.
The bare form factors $\hat F^I(\hat a_s,Q^2,\mu^2,\ep)$
corresponding to the unrenormalised operators satisfy
the Sudakov differential equation \cite{Sudakov:1954sw,Mueller:1979ih,
Collins:1980ih,Sen:1981sd}.  In dimensional regularisation,
the formal solution to the differential equation
up to four loop level can be found in \cite{Moch:2005id,Ravindran:2005vv}
in terms of  $A^I$, the standard cusp anomalous dimensions,
and the constants $G^{~I}_i(\ep)$ for both $I=q,b$ and
$I=g$ \cite{Moch:2005tm} to the required accuracy in $\ep$.
The single poles of the form factors contain
the combination
\cite{Moch:2005id,Ravindran:2006cg,Ravindran:2004mb,Ravindran:2005vv}
$$2 \Bigg(B^{~I}_i - \delta_{I,g}~ i~\beta_{i-1} -\delta_{I,b}
\gamma^b_{i-1}\Bigg) + f^{~I}_i$$ at order $\hat a_s^i$.
The terms proportional $-2(\delta_{I,g}~i~ \beta_{i-1} +\delta_{I,b}
\gamma^b_{i-1})$ come from the large momentum region of the loop integrals
that are giving ultraviolet divergences.  The poles containing them
will go away when the form factors undergo overall operator
UV renormalisation through the renormalisation constants
$Z^{~I}$ which satisfy the RG equations
\begin{eqnarray}
\mu_R^2 {d \over d\mu_R^2} \ln Z^g(\hat a_s,\mu_R^2,\mu^2,\ep) &=&
\sum_{i=1}^\infty a^i_s(\mu_R^2)~ \Big(i~\beta_{i-1}\Big)
\nonumber\\
\mu_R^2 {d \over d\mu_R^2}\ln Z^b(\hat a_s,\mu_R^2,\mu^2,\ep)&=&
\sum_{i=1}^\infty a^i_s(\mu_R^2)~ \gamma^b_{i-1}
\end{eqnarray}
where $\ep \rightarrow 0$ is set.  The constants
$i~\beta_{i-1}$ and $\gamma^b_{i-1}$ are anomalous dimensions
of the renormalised form factors $F^g$ and $F^b$ respectively.
After the overall operator renormalisation through $Z^{~I}$ and coupling
constant renormalisation through $Z$, the remaining poles will be
proportional to  $B^{~I}_i$ and $f^{~I}_i$ in addition to the
standard cusp anomalous dimensions $A^{~I}_i$.
The constants $B_i^I$ and $f_i^I$  are
known up to order $a_s^3$\cite {Moch:2004pa,Vogt:2004mw,Ravindran:2004mb,Moch:2005tm}.

The collinear singularities resulting from massless partons are
removed in $\overline {MS}$ scheme using
the mass factorisation kernel $\widetilde\Gamma(z,\mu_F^2,\ep)$.
The kernel $\widetilde \Gamma(z,\mu_F^2,\ep)$ satisfies the
following renormalisation group equation:
\begin{eqnarray}
\mu_F^2 {d \over d\mu_F^2}
\widetilde \Gamma(z,\mu_F^2,\ep)={1 \over 2}  \widetilde P
                         \left(z,\mu_F^2\right)
                        \otimes
               \widetilde \Gamma \left(z,\mu_F^2,\ep\right)~.
\end{eqnarray}
The function $\widetilde P(z,\mu_F^2)$ are the well known 
(matrix-valued) Altarelli-Parisi 
splitting
functions 
\begin{eqnarray}
\widetilde P(z,\mu_F^2)=
\sum_{i=1}^{\infty}a_s^i(\mu_F^2) \widetilde P^{(i-1)}(z)
\end{eqnarray}
The diagonal terms of the splitting functions
$\widetilde P^{(i)}(z)$ have the following structure
\begin{eqnarray}
\widetilde P^{(i)}_{II}(z) = 2\Bigg[ B^I_{i+1} \delta(1-z)
                  + A^I_{i+1} {\cal D}_0\Bigg] 
+ \widetilde P_{reg,II}^{(i)}(z)~,
\end{eqnarray}
where $\widetilde P_{reg,II}^{(i)}$ are regular
when $z \rightarrow 1$.
In the case of the soft plus virtual part of the decay distributions,
only the diagonal parts of the kernels contribute.  In the $\overline{MS}$ scheme,
the kernel contains only poles in $\ep$.  The kernel can be expanded in powers
of bare coupling $\hat a_s$ as
\begin{eqnarray}
\widetilde \Gamma(z,\mu_F^2,\ep)=\delta(1-z)+\sum_{i=1}^\infty \hat a_s^i
\left({\mu_F^2 \over \mu^2}\right)^{i {\ep \over 2}}S^i_{\ep}
\widetilde \Gamma^{(i)}(z,\ep)~.
\end{eqnarray}
The constants $\widetilde \Gamma^{(i)}(z,\ep)$ are expanded in negative 
powers of $\ep$ up to the four loop level can be read from 
Ref.~\cite{Ravindran:2005vv}
where similar RG equations were solved.

It is natural to expect that the soft plus jet
distribution functions have a pole structure in $\ep$ similar to that
of $\hat F^I$ and $\widetilde \Gamma_{II}$ so that
the decay distributions $\widetilde \Delta^{sv}_{~I}$ are finite in the limit
$\ep \rightarrow 0$.  This implies that they satisfy a Sudakov type
differential equation that
the form factors $\hat F^I$ do.
Solving the Sudakov differential equation for 
$\widetilde \Phi^{~I}(\hat a_s,q^2,\mu^2,z,\ep)$, we get
\begin{eqnarray}
\widetilde \Phi^{~I}(\hat a_s,q^2,\mu^2,z,\ep) &=&
\widetilde \Phi^{~I}(\hat a_s,q^2 (1-z),\mu^2,\ep)
\nonumber\\[2ex]
&=&\sum_{i=1}^\infty \hat a_s^i \left({q^2 (1-z)
\over \mu^2}\right)^{i {\ep \over 2}} S_{\ep}^i
\left({i~ \ep \over 2(1-z)} \right)\hat \phi^{~I,(i)}(\ep)
\label{SpJ}
\end{eqnarray}
where $\hat \phi^{~I,(i)}(\ep)$ coincide with the $\hat \phi_{SJ}^{~I,(i)}(\ep)$
that appear in the deep-inelastic scattering cross 
section\cite{Ravindran:2006cg}.
This is the result
of exact crossing symmetry between the processes under study and the DIS
process in the threshold region.
The constants $\hat \phi_{SJ}^{~I,(i)}(\ep)$ for $I=q$ corresponding
to DIS are given in \cite{Ravindran:2005vv,Ravindran:2006cg}.  Since 
the $b$ quark is treated massless,
we use the same constants for the Higgs boson decay also.

The threshold corrections that dominate when the partonic scaling
variable $z$ approaches its kinematic limit, which is $1$, through
the distributions $\delta(1-z)$ and ${\cal D}_i$ can easily be resummed
in Mellin $N$ because the decay distributions appear as convolution
of partonic distributions and fragmentation functions.
This has been a successful
approach due to several important works
(see \cite{Kodaira:1981nh,Sterman:1986aj,Catani:1989ne,Vogt:2000ci,Cacciari:2001cw}).
The higher order threshold exponents can be found in
\cite{Moch:2005ba,Laenen:2005uz,Idilbi:2005ni,Ravindran:2005vv}.
We find that the soft plus jet distribution function
$\widetilde \Phi^I(\hat a_s,q^2,\mu^2,z,\ep)$
captures all the features of the $N$ space resummation approach by
expressing (\ref{SpJ}) as
\begin{eqnarray}
\widetilde \Phi^{~I}(\hat a_s,q^2,\mu^2,z,\ep)&=&
\Bigg( {1 \over 2 (1-z)} \Bigg\{
\int_{\mu_R^2}^{q^2 (1-z)} {d \lambda^2 \over \lambda^2}
A_I \left(a_s(\lambda^2)\right) + \overline G^{~I}_{SJ} \left(
a_s\left(q^2 (1-z)\right),\ep\right)\Bigg\} \Bigg)_+
\nonumber\\[2ex]
&&+\delta(1-z)~~~ \sum_{i=1}^\infty \hat a_s^i
\left({q^2 \over \mu^2}\right)^{i {\ep \over 2}}
S_{\ep}^i~
\hat \phi^{~I,(i)}_{SJ}(\ep)
\nonumber\\[2ex]
&&+\left({1 \over 2(1-z)}\right)_+ ~~~\sum_{i=1}^\infty
\hat a_s^i \left({\mu_R^2 \over \mu^2}\right)^{i {\ep \over 2}}
S_{\ep}^i~
\overline K^{~I,(i)}(\ep)
\label{resum}
\end{eqnarray}
where
\begin{eqnarray}
\overline G^{~I}_{SJ}\left(a_s \left(q^2 (1-z) \right),\ep\right)
&=& \sum_{i=1}^\infty \hat a_s^i
\left({q^2 (1-z) \over \mu^2}\right)^{i{\ep \over 2}}
S_{\ep}^i
\overline G^{~I,(i)}_{SJ}(\ep)~.
\label{Gbar2}
\end{eqnarray}
Here, the constants $\overline G^{~I,(i)}_{SJ}(\ep),\overline 
K^{~I,(i)}(\ep)$
and $\hat \phi^{~I,(i)}_{SJ}(\ep)$
can be found in \cite{Ravindran:2005vv,Ravindran:2006cg}.
The second line of the equation (\ref{resum}) contains the right poles
in $\ep$ to cancel those coming from the form factor as well as
from the mass factorisation kernel.  It also
contains the terms that are finite as $\ep$ becomes zero through
the constants $\overline {\cal G}^{~I}_{JS,i}(\ep)$ that appear in
$\overline G^{~I,(i)}_{SJ}(\ep)$ (see \cite{Ravindran:2006cg}).
Since they are multiplied by $\delta(1-z)$,
the fixed order soft plus virtual part of the decay distributions
gets contributions from them at higher orders.
The third line in the eqn.(\ref{resum}) contains only poles in $\ep$ and
they all cancel against ${\cal D}_0$ parts of the mass factorisation kernel.
Hence, adding the eqn.(\ref{resum}) with the renormalised form factor and the
mass factorisation kernel and performing the coupling constant renormalisation
and then finally taking Mellin moment,
we reproduce the exponents in resummation formula(see
\cite{Kodaira:1981nh,Sterman:1986aj,Catani:1989ne,Vogt:2000ci,Cacciari:2001cw})
(after setting $\ep \rightarrow 0$).
We find that the function
$\overline G^{~I}_{SJ}\left(a_s \left(q^2 (1-z) \right),\ep\right)$
appearing in the first line of eqn.(\ref{resum})
coincide with the exponent
$B^I_{decay}\left(a_s\left(q^2 (1-z)\right)\right)$ appearing in the resummation
formula
\begin{eqnarray}
B^I_{decay}\left(a_s\left(q^2 (1-z)\right)\right)
&=&\sum_{i=1}^\infty a_s^i\left(q^2 (1-z) \right) B_{decay,i}^I
\nonumber\\[2ex]
&=&\overline G^{~I}_{SJ}\left(a_s \left(q^2 (1-z) \right),\ep\right)
\Bigg|_{\ep=0}~.
\end{eqnarray}
In addition, we get a new exponent
that comes from the Mellin moment of $\delta(1-z)$ part and hence
$N$ independent.  The Mellin space resummation exponents contain
the cusp anomalous dimensions $A^I$ and the constants $B_{decay,i}^I$
which we find are identical to those of DIS.  These constants are known
up to three loop level whose numerical impacts on the physical observables
will be important in order to reduce theoretical uncertainties coming from
renormalisation and factorisation scales.

Using the $z$ space resummed expression given in Eq.~(\ref{master}) and 
the 
known
exponents, we present here the results for $\widetilde \Delta_{I}^{sv,(i)}$ for
the bottom quark energy distribution in Higgs boson decay beyond leading 
order
in QCD.  We present the complete soft plus virtual contribution to order 
$N^iLO$
for $i=1,2$ and a partial $N^3LO$ result, i.e., without $\delta(1-z)$ 
part.
In addition, we extend this approach to $N^4LO$ order where we can obtain
the {\it partial} soft plus virtual contribution coming
from all ${\cal D}_j$ except $j=0,1$
for the $N^4LO$ coefficient $\widetilde \Delta_{b}^{sv,(4)}$.
Here also we cannot determine the $\delta(1-z)$ part.
We repeat this for the hadroproduction in $l^+l^-$ annihilation
up to $N^4LO$ order.  These fixed order results are expected to reduce
the renormalisation and factorisation scale uncertainties.
Expanding
$\widetilde \Delta_{I}^{sv}(z,q^2,\mu_R^2,\mu_F^2)$ in powers of
$a_s(\mu_R^2)$ as
\begin{eqnarray}
\widetilde \Delta_{I}^{sv}(z,q^2,\mu_R^2,\mu_F^2)&=&\sum_{i=0}^\infty a_s^i(\mu_R^2)~
\widetilde \Delta_{I}^{sv,(i)}(z,q^2,\mu_R^2,\mu_F^2)
\end{eqnarray}
and choosing $\mu_R^2=\mu_F^2=q^2$ for simplicity, we find
\begin{eqnarray}
   \widetilde \Delta_b^{sv,(0)}& = &\delta(1-z)
\\
   \widetilde \Delta_b^{sv,(1)}& = &
        \delta(1-z)~
\Bigg[C_F ~ \Big(
           3
          + 8~\zeta_2
          \Big)
\Bigg]
       + {\cal D}_0~
\Bigg[C_F ~ \Big(
          - 3
          \Big)
\Bigg]
       + {\cal D}_1~
\Bigg[C_F ~ \Big(
           4
          \Big)
\Bigg]
\\
   \widetilde \Delta_b^{sv,(2)} &=&
        \delta(1-z)~
\Bigg[n_f~C_F~  \Big(
          - 91/12
          - 14/3~\zeta_2
          + 4/3~\zeta_3
          \Big)
       + C_F~C_A~  \Big(
           1691/24
          + 95/3~\zeta_2
\nonumber\\&&
          - 49/5~\zeta_2^2
          + 32/3~\zeta_3
          \Big)
       + C_F^2  ~\Big(
           109/8
          + 31~\zeta_2
          + 30~\zeta_2^2
          - 78~\zeta_3
          \Big)
\Bigg]
\nonumber\\&&
       + {\cal D}_0~
\Bigg[n_f~C_F ~ \Big(
           247/27
          - 8/3~\zeta_2
          \Big)
       + C_F~C_A ~ \Big(
          - 3155/54
          + 44/3~\zeta_2
          + 40~\zeta_3
          \Big)
\nonumber\\&&
       + C_F^2 ~ \Big(
          - 21/2
          - 8~\zeta_3
          \Big)
\Bigg]
       + {\cal D}_1~
\Bigg[n_f~C_F ~ \Big(
          - 58/9
          \Big)
       + C_F~C_A ~ \Big(
           367/9
          - 8~\zeta_2
          \Big)
\nonumber\\&&
       + C_F^2 ~ \Big(
           21
          + 16~\zeta_2
          \Big)
\Bigg]
       + {\cal D}_2~
\Bigg[n_f~C_F ~ \Big(
           4/3
          \Big)
       + C_F~C_A ~ \Big(
          - 22/3
          \Big)
       + C_F^2 ~ \Big(
          - 18
          \Big)
\Bigg]
\nonumber\\&&
       + {\cal D}_3~
\Bigg[C_F^2 ~ \Big(
           8
          \Big)
\Bigg]
\\
   \widetilde \Delta_b^{sv,(3)} &=&
        {\cal D}_0~
\Bigg[n_f~C_F~C_A ~ \Big(
           160906/729
          - 9920/81~\zeta_2
          + 208/15~\zeta_2^2
          - 776/9~\zeta_3
          \Big)
\nonumber\\&&
       + n_f~C_F^2 ~ \Big(
           16423/108
          - 722/27~\zeta_2
          - 16~\zeta_2^2
          - 60~\zeta_3
          \Big)
       + n_f^2~C_F ~ \Big(
          - 8714/729
\nonumber\\&&
          + 232/27~\zeta_2
          - 32/27~\zeta_3
          \Big)
       + C_F~C_A^2 ~ \Big(
          - 599375/729
          - 176/3~\zeta_2~\zeta_3
          + 32126/81~\zeta_2
\nonumber\\&&
          - 652/15~\zeta_2^2
          + 21032/27~\zeta_3
          - 232~\zeta_5
          \Big)
       + C_F^2~C_A ~ \Big(
          - 31151/72
          + 80~\zeta_2~\zeta_3
\nonumber\\&&
          + 3365/27~\zeta_2
          + 39~\zeta_2^2
          + 1988/9~\zeta_3
          - 120~\zeta_5
          \Big)
       + C_F^3 ~ \Big(
          - 479/8
          - 64~\zeta_2~\zeta_3
\nonumber\\&&
          - 45~\zeta_2
          + 6~\zeta_2^2
          + 178~\zeta_3
          + 432~\zeta_5
          \Big)
\Bigg]
       + {\cal D}_1~
\Bigg[n_f~C_F~C_A ~ \Big(
          - 15062/81
          + 512/9~\zeta_2
\nonumber\\&&
          + 16~\zeta_3
          \Big)
       + n_f~C_F^2 ~ \Big(
          - 1343/9
          + 64/3~\zeta_2
          + 112/3~\zeta_3
          \Big)
       + n_f^2~C_F ~ \Big(
           940/81
          - 32/9~\zeta_2
          \Big)
\nonumber\\&&
       + C_F~C_A^2 ~ \Big(
           50689/81
          - 680/3~\zeta_2
          + 176/5~\zeta_2^2
          - 264~\zeta_3
          \Big)
       + C_F^2~C_A ~ \Big(
           13783/18
\nonumber\\&&
          - 352/3~\zeta_2
          - 196/5~\zeta_2^2
          - 592/3~\zeta_3
          \Big)
       + C_F^3 ~ \Big(
           181/2
          + 4~\zeta_2
          - 104/5~\zeta_2^2
          - 360~\zeta_3
          \Big)
\Bigg]
\nonumber\\&&
       + {\cal D}_2~
\Bigg[n_f~C_F~C_A ~ \Big(
           1552/27
          - 16/3~\zeta_2
          \Big)
       + n_f~C_F^2 ~ \Big(
           827/9
          - 64/3~\zeta_2
          \Big)
\nonumber\\&&
       + n_f^2~C_F ~ \Big(
          - 116/27
          \Big)
       + C_F~C_A^2 ~ \Big(
          - 4649/27
          + 88/3~\zeta_2
          \Big)
       + C_F^2~C_A ~ \Big(
          - 10009/18
\nonumber\\&&
          + 460/3~\zeta_2
          + 240~\zeta_3
          \Big)
       + C_F^3 ~ \Big(
          - 153/2
          + 72~\zeta_2
          + 16~\zeta_3
          \Big)
\Bigg]
\nonumber\\&&
       + {\cal D}_3~
\Bigg[n_f~C_F~C_A ~ \Big(
          - 176/27
          \Big)
       + n_f~C_F^2 ~ \Big(
          - 280/9
          \Big)
       + n_f^2~C_F ~ \Big(
           16/27
          \Big)
\nonumber\\&&
       + C_F~C_A^2 ~ \Big(
           484/27
          \Big)
       + C_F^2~C_A ~ \Big(
           1732/9
          - 32~\zeta_2
          \Big)
       + C_F^3 ~ \Big(
           60
          \Big)
\Bigg]
       + {\cal D}_4~
\Bigg[n_f~C_F^2 ~ \Big(
           40/9
          \Big)
\nonumber\\&&
       + C_F^2~C_A ~ \Big(
          - 220/9
          \Big)
       + C_F^3 ~ \Big(
          - 30
          \Big)
\Bigg]
       + {\cal D}_5~
\Bigg[C_F^3 ~ \Big(
           8
          \Big)
\Bigg]
\\
   \widetilde \Delta_b^{sv,(4)}&=&
        {\cal D}_2~
\Bigg[n_f~C_F~C_A^2 ~ \Big(
           17189/9
          - 5096/9~\zeta_2
          + 176/5~\zeta_2^2
          - 352~\zeta_3
          \Big)
\nonumber\\&&
       + n_f~C_F^2~C_A ~ \Big(
           964334/243
          - 57524/27~\zeta_2
          + 2332/15~\zeta_2^2
          - 11032/9~\zeta_3
          \Big)
\nonumber\\&&
       + n_f~C_F^3 ~ \Big(
           12299/9
          - 2336/3~\zeta_2
          - 728/5~\zeta_2^2
          - 936~\zeta_3
          \Big)
       + n_f^2~C_F~C_A ~ \Big(
          - 7403/27
\nonumber\\&&
          + 688/9~\zeta_2
          + 16~\zeta_3
          \Big)
       + n_f^2~C_F^2 ~ \Big(
          - 71776/243
          + 4088/27~\zeta_2
          + 304/9~\zeta_3
          \Big)
\nonumber\\&&
       + n_f^3~C_F ~ \Big(
           940/81
          - 32/9~\zeta_2
          \Big)
       + C_F~C_A^3 ~ \Big(
          - 649589/162
          + 4012/3~\zeta_2
          - 968/5~\zeta_2^2
\nonumber\\&&
          + 1452~\zeta_3
          \Big)
       + C_F^2~C_A^2 ~ \Big(
          - 6034493/486
          - 832~\zeta_2~\zeta_3
          + 63764/9~\zeta_2
          - 2450/3~\zeta_2^2
\nonumber\\&&
          + 25336/3~\zeta_3
          - 1392~\zeta_5
          \Big)
       + C_F^3~C_A ~ \Big(
          - 12529/3
          - 768~\zeta_2~\zeta_3
          + 13718/3~\zeta_2
\nonumber\\&&
          + 1094/5~\zeta_2^2
          + 13300/3~\zeta_3
          - 720~\zeta_5
          \Big)
       + C_F^4 ~ \Big(
          - 420
          - 448~\zeta_2~\zeta_3
\nonumber\\&&
          + 198~\zeta_2
          + 324~\zeta_2^2
          + 1584~\zeta_3
          + 4128~\zeta_5
          \Big)
\Bigg]
\nonumber\\&&
       + {\cal D}_3~
\Bigg[n_f~C_F~C_A^2 ~ \Big(
          - 9502/27
          + 352/9~\zeta_2
          \Big)
       + n_f~C_F^2~C_A ~ \Big(
          - 358142/243
\nonumber\\&&
          + 11984/27~\zeta_2
          + 1216/9~\zeta_3
          \Big)
       + n_f~C_F^3 ~ \Big(
          - 5660/9
          + 2608/9~\zeta_2
          + 1888/9~\zeta_3
          \Big)
\nonumber\\&&
       + n_f^2~C_F~C_A ~ \Big(
           1540/27
          - 32/9~\zeta_2
          \Big)
       + n_f^2~C_F^2 ~ \Big(
           25966/243
          - 736/27~\zeta_2
          \Big)
\nonumber\\&&
       + n_f^3~C_F ~ \Big(
          - 232/81
          \Big)
       + C_F~C_A^3 ~ \Big(
           55627/81
          - 968/9~\zeta_2
          \Big)
       + C_F^2~C_A^2 ~ \Big(
           2259107/486
\nonumber\\&&
          - 47104/27~\zeta_2
          + 864/5~\zeta_2^2
          - 13024/9~\zeta_3
          \Big)
       + C_F^3~C_A ~ \Big(
           26720/9
          - 16192/9~\zeta_2
\nonumber\\&&
          + 248/5~\zeta_2^2
          - 11392/9~\zeta_3
          \Big)
       + C_F^4 ~ \Big(
           533/2
          - 376~\zeta_2
          - 1488/5~\zeta_2^2
          - 1072~\zeta_3
          \Big)
\Bigg]
\nonumber\\&&
       + {\cal D}_4~
\Bigg[n_f~C_F~C_A^2 ~ \Big(
           242/9
          \Big)
       + n_f~C_F^2~C_A ~ \Big(
           8120/27
          - 80/3~\zeta_2
          \Big)
       + n_f~C_F^3 ~ \Big(
           6070/27
\nonumber\\&&
          - 560/9~\zeta_2
          \Big)
       + n_f^2~C_F~C_A ~ \Big(
          - 44/9
          \Big)
       + n_f^2~C_F^2 ~ \Big(
          - 640/27
          \Big)
       + n_f^3~C_F ~ \Big(
           8/27
          \Big)
\nonumber\\&&
       + C_F~C_A^3 ~ \Big(
          - 1331/27
          \Big)
       + C_F^2~C_A^2 ~ \Big(
          - 24040/27
          + 440/3~\zeta_2
          \Big)
       + C_F^3~C_A ~ \Big(
          - 35755/27
\nonumber\\&&
          + 4160/9~\zeta_2
          + 400~\zeta_3
          \Big)
       + C_F^4 ~ \Big(
          - 150
          + 240~\zeta_2
          + 400/3~\zeta_3
          \Big)
\Bigg]
\nonumber\\&&
       + {\cal D}_5~
\Bigg[n_f~C_F^2~C_A ~ \Big(
          - 704/27
          \Big)
       + n_f~C_F^3 ~ \Big(
          - 164/3
          \Big)
       + n_f^2~C_F^2 ~ \Big(
           64/27
          \Big)
\nonumber\\&&
       + C_F^2~C_A^2 ~ \Big(
           1936/27
          \Big)
       + C_F^3~C_A ~ \Big(
           998/3
          - 48~\zeta_2
          \Big)
       + C_F^4 ~ \Big(
           78
          - 32~\zeta_2
          \Big)
\Bigg]
\nonumber\\&&
       + {\cal D}_6~
\Bigg[n_f~C_F^3 ~ \Big(
           56/9
          \Big)
       + C_F^3~C_A ~ \Big(
          - 308/9
          \Big)
       + C_F^4 ~ \Big(
          - 28
          \Big)
\Bigg]
       + {\cal D}_7~
\Bigg[C_F^4 ~ \Big(
           16/3
          \Big)
\Bigg]
\end{eqnarray}
Similarly for  hadroproduction in $l^+l^-$ annihilation, we find
\begin{eqnarray}
\widetilde \Delta_b^{sv,(0)} -\widetilde \Delta_q^{sv,(0)}&=&0
\\
\widetilde \Delta_b^{sv,(1)} -\widetilde \Delta_q^{sv,(1)}&=&
\delta(1-z)\Bigg[ C_F   \Big(
           12
          \Big)\Bigg]
\\
\widetilde \Delta_b^{sv,(2)} -\widetilde \Delta_q^{sv,(2)}&=&
        \delta(1-z)
\Bigg[n_f C_F   \Big(
          - 365/18
          + 8 ~\zeta_2
          \Big)
       + C_F C_A   \Big(
           5269/36
          - 40 ~\zeta_2
          - 36 ~\zeta_3
          \Big)
\nonumber\\&&
       + C_F^2   \Big(
          - 111/4
          + 70 ~\zeta_2
          \Big)
\Bigg]
       + {\cal D}_0
\Bigg[C_F^2   \Big(
          - 36
          \Big)
\Bigg]
       + {\cal D}_1
\Bigg[C_F^2   \Big(
           48
          \Big)
\Bigg]
\\
   \widetilde \Delta_b^{sv,(3)} -\widetilde \Delta_q^{sv,(3)}&=&
        {\cal D}_0
\Bigg[n_f C_F^2   \Big(
           3071/18
          - 56 ~\zeta_2
          \Big)
       + C_F^2 C_A   \Big(
          - 41047/36
          + 296 ~\zeta_2
          + 588 ~\zeta_3
          \Big)
\nonumber\\&&
       + C_F^3   \Big(
           261/4
          + 78 ~\zeta_2
          - 96 ~\zeta_3
          \Big)
\Bigg]
       + {\cal D}_1
\Bigg[n_f C_F^2   \Big(
          - 1426/9
          + 32 ~\zeta_2
          \Big)
\nonumber\\&&
       + C_F^2 C_A   \Big(
           9673/9
          - 256 ~\zeta_2
          - 144 ~\zeta_3
          \Big)
       + C_F^3   \Big(
          - 3
          + 88 ~\zeta_2
          \Big)
\Bigg]
\nonumber\\&&
       + {\cal D}_2
\Bigg[n_f C_F^2   \Big(
           16
          \Big)
       + C_F^2 C_A   \Big(
          - 88
          \Big)
       + C_F^3   \Big(
          - 216
          \Big)
\Bigg]
       + {\cal D}_3
\Bigg[C_F^3   \Big(
           96
          \Big)
\Bigg]
\\
   \widetilde \Delta_b^{sv,(4)} -\widetilde \Delta_q^{sv,(4)}&=&
        {\cal D}_2
\Bigg[n_f C_F^2 C_A   \Big(
           27908/27
          - 176 ~\zeta_2
          - 48 ~\zeta_3
          \Big)
       + n_f C_F^3   \Big(
           4148/3
          - 1304/3 ~\zeta_2
          \Big)
\nonumber\\&&
       + n_f^2 C_F^2   \Big(
          - 2122/27
          + 32/3 ~\zeta_2
          \Big)
       + C_F^2 C_A^2   \Big(
          - 169535/54
          + 1936/3 ~\zeta_2
\nonumber\\&&
          + 264 ~\zeta_3
          \Big)
       + C_F^3 C_A   \Big(
          - 26519/3
          + 8252/3 ~\zeta_2
          + 3528 ~\zeta_3
          \Big)
       + C_F^4   \Big(
           459/2
\nonumber\\&&
          + 1332 ~\zeta_2
          + 192 ~\zeta_3
          \Big)
\Bigg]
       + {\cal D}_3
\Bigg[n_f C_F^2 C_A   \Big(
          - 704/9
          \Big)
       + n_f C_F^3   \Big(
          - 4820/9
\nonumber\\&&
          + 64 ~\zeta_2
          \Big)
       + n_f^2 C_F^2   \Big(
           64/9
          \Big)
       + C_F^2 C_A^2   \Big(
           1936/9
          \Big)
       + C_F^3 C_A   \Big(
           31322/9
\nonumber\\&&
          - 704 ~\zeta_2
          - 288 ~\zeta_3
          \Big)
       + C_F^4   \Big(
           210
          - 208 ~\zeta_2
          \Big)
\Bigg]
       + {\cal D}_4
\Bigg[n_f C_F^3   \Big(
           160/3
          \Big)
\nonumber\\&&
       + C_F^3 C_A   \Big(
          - 880/3
          \Big)
       + C_F^4   \Big(
          - 360
          \Big)
\Bigg]
       + {\cal D}_5
\Bigg[C_F^4   \Big(
           96
          \Big)~,
\Bigg]
\end{eqnarray}
where the colour factors for the $SU(N)$ gauge group are 
$C_A=N,C_F=(N^2-1)/(2N)$ and
$n_f$ is the number of active flavours. The above expressions are easily 
transformed to Mellin space. The functions ${\cal D}_k$ are represented
by polynomials of single harmonic sums only, cf. \cite{BK}. The soft 
resummation terms supplement the corresponding representations for 
the 2--loop Wilson coefficients of the production cross sections in Mellin 
space derived in \cite{BR}.  

To summarise, we have systematically studied the soft plus virtual
contributions to the bottom quark energy distribution in Higgs boson decay
and the hadroproductions in $l^+l^-$ annihilation processes
using the formalism of threshold resummation to infrared safe decay distributions.
This was achieved using renormalisation group invariance and
Sudakov resummation of soft gluons and the factorisation
property of these decay distributions.
We have also shown how these resummed distributions
are related to resummation exponents that appear in Mellin space.
Using this approach we have computed the soft plus virtual decay 
distributions
at $NNLO$ and partial results at $N^3LO$ and $N^4LO$ in perturbative QCD
for the bottom quark energy distribution in Higgs decay and 
hadroproductions
in $l^+l^-$ annihilation.

\vspace{2mm}\noindent
{\bf Acknowledgments:}  VR would like to thank Prakash Mathews for his 
constant
encouragements and discussions. This work was supported in part by DFG 
Sonderforschungsbereich Transregio 9, Computergest\"utzte Theoretische 
Physik.


\end{document}